\documentclass[twocolumn]{aastex61}

\usepackage[latin1]{inputenc}
\usepackage{epstopdf}
\usepackage{amssymb}
\usepackage{amsmath}
\usepackage{color}

\usepackage{rotating}
\usepackage{multirow}
\def\kms{km\,s$^{-1}$}
\def\Ha{H{$\alpha$}}
\def\M{M$_{\odot}$}

\newcommand{\bluemag}[1]{\ensuremath{M(400)_\mathrm{#1}}}
\newcommand{\redmag}[1]{\ensuremath{M(520)_\mathrm{#1}}}
\newcommand{\peakcol}{\ensuremath{\bluemag{0}-\redmag{0}}}
\newcommand{\thirtycol}{\ensuremath{\bluemag{30}-\redmag{30}}}

\shorttitle{SLSN identification and diversity}
\shortauthors{Inserra et al.}

\begin{document}

\title{A statistical approach to identify superluminous supernovae and probe their diversity}

\author{C. Inserra}
\affiliation{Department of Physics and Astronomy, University of Southampton, Southampton, SO17 1BJ, UK}

\author{S. Prajs}
\affiliation{Department of Physics and Astronomy, University of Southampton, Southampton, SO17 1BJ, UK}

\author{C. P. Gutierrez}
\affiliation{Department of Physics and Astronomy, University of Southampton, Southampton, SO17 1BJ, UK}

\author{C. Angus}
\affiliation{Department of Physics and Astronomy, University of Southampton, Southampton, SO17 1BJ, UK}

\author{M. Smith}
\affiliation{Department of Physics and Astronomy, University of Southampton, Southampton, SO17 1BJ, UK}

\author{M. Sullivan}
\affiliation{Department of Physics and Astronomy, University of Southampton, Southampton, SO17 1BJ, UK}

%\author{
%C. Inserra\altaffilmark{1}, 
%S. Prajs\altaffilmark{1},
%G. P. Gutierrez\altaffilmark{1},
%C. Angus\altaffilmark{1},
%M. Smith\altaffilmark{1} and
%M. Sullivan\altaffilmark{1}.
%%plus others
%%FINAL ORDER TBD
%}
%
%\altaffiltext{1}{Department of Physics \& Astronomy, University of Southampton, Southampton, Hampshire, SO17 %1BJ, UK; C.Inserra@soton.ac.uk}
%%\altaffiltext{2}{Astrophysics Research Centre, School of Mathematics and Physics, Queens University
% % Belfast, Belfast BT7 1NN, UK}

\email{C.Inserra@soton.ac.uk}

\begin{abstract}
We investigate the identification of hydrogen-poor superluminous supernovae (SLSNe~I) using a photometric analysis, without including an arbitrary magnitude threshold. We assemble a homogeneous sample of previously classified SLSNe~I from the literature, and fit their light curves using Gaussian processes. From the fits, we identify four photometric parameters that have a high statistical significance when correlated, and combine them in a parameter space that conveys information on their luminosity and color evolution. 
This parameter space presents a new  definition for SLSNe~I, which can be used to analyse existing and future transient datasets. We find that 90\% of previously classified SLSNe~I meet our new definition. We also examine the evidence for two subclasses of SLSNe~I, combining their photometric evolution with spectroscopic information, namely the photospheric velocity and its gradient.
A cluster analysis reveals the presence of two distinct groups. `Fast' SLSNe show fast light curves and color evolution, large velocities, and a large velocity gradient. `Slow' SLSNe show slow light curve and color evolution, small expansion velocities, and an almost non-existent velocity gradient. Finally, we discuss the impact of our analyses in the understanding of the powering engine of SLSNe, and their implementation as cosmological probes in current and future surveys.
\end{abstract}


\section{Introduction}
\label{sec:intro}

The last decade of observations by untargeted optical time-domain surveys has unveiled a population of exceptionally bright optical transients, with peak magnitudes of M~$\lesssim-21$, labeled ``superluminous supernovae" \citep[SLSNe;][]{qu11,gy12}. There are two broad classes: SLSNe~II, which exhibit signatures of hydrogen in their optical spectra, and SLSNe~I (or SLSNe~Ic), which do not. SLSNe~II are heterogeneous in both luminosity and host environment \citep[][]{gy09,le15,sc16}, and the bulk of the population consists of events displaying signatures of interaction similar to classical SNe IIn \citep[e.g., SN2006gy;][]{smi07}, with a smaller contribution from intrinsically bright events reminiscent of classical SNe II \citep[e.g., SNe 2008es, 2013hx;][]{mi09,ge09,in16}.
Hydrogen-poor SLSNe \citep[][]{qu11,gy12,in13} are the most common type of SLSN, although they are still intrinsically rare compared to other SN types \citep[91$^{+76}_{-36}$ SNe yr$^{-1}$ Gpc$^{-3}$,][]{pr17}, and spectroscopically linked to normal or broad-lined type Ic SNe  \citep{pasto10}. Their characteristic spectroscopic evolution and connection with massive star explosions have been a distinctive trait of SLSNe~I, together with their typical explosion location in dwarf, metal-poor and star-forming galaxies \citep[e.g.][]{lu14,le15,an16,pe16,ch17}.

This relatively simple description and overall SLSN~I paradigm is now becoming more complex. The original definition of SLSNe~I has been loosened in terms of the magnitude threshold \citep[e.g.,][]{in13,lu16,pr17},
and the existence of two subclasses of SLSNe~I, with a difference in the speed of their light-curve evolution (slow- versus fast-evolving), is debated. The concept that all SLSNe~I originate from the same progenitor scenario and/or explosion mechanism, with differences principally driven by variations in ejecta mass \citep{ni15}, has been challenged by their spectroscopic evolution \citep{in17}. 

This complexity is increasing with new data releases from the Palomar Transient Factory \citep{2017arXiv170801623D} and the Pan-STARRS1 Medium Deep Survey \citep{lu17}, and with future data releases from the Dark Energy Survey (DES) Supernova Program \citep[DES-SN;][]{dessn}. Moreover, the next generation of telescopes will likely bring an order of magnitude increase in sample sizes: \citet{sco16} predicted 10,000 SLSNe~I will be discovered by the Large Synoptic Survey Telescope (LSST), and \citet{in17b} calculate a discovery rate of $\sim$200 SLSNe~I during the five-year deep field survey of the European Space Agency {\it Euclid} satellite, with the potential for further discoveries up to $z\sim6$ \citep[e.g.][]{yan17,smi17} with the \textit{Wide-Field InfraRed Survey Telescope} (\textit{WFIRST}). Mapping the spectroscopic evolution of such a large number of targets will be challenging to achieve.

In this paper, we introduce a new statistical approach to defining and classifying SLSNe that could be used in current (e.g., DES, PTF, and Pan-STARRS1) and forthcoming large samples of SLSN~I candidates. The technique neither assumes an arbitrary magnitude limit, nor relies on a detailed spectroscopic evolution, and is developed using the published data set of SLSNe~I. We also present an analysis, based on statistical tools, showing the existence of two subclasses from their spectrophotometric evolution. These two methodologies, combined together, will characterize and classify a homogeneous sample of SLSNe~I, as well as select those events that can be used as cosmological standardizable candles \citep[see][]{in14}.

\section{Sample and methodology}\label{sec:sgp}

\subsection{Sample selection}\label{ss:sample}
\label{sec:sample}

We begin by constructing our SLSN~I sample from published events in the literature. We require a well-observed sample, with at least six epochs of photometric data between phases of $-15$\,d to $+30$\,d, of which at least one must lie between $-15$\,d and $0$\,d, sampling a synthetic box filter at rest-frame 4000\,\AA.
This synthetic filter was introduced in \citet{in14}, together with a second box filter at 5200\,\AA, which are designed to sample two regions of SLSN~I spectra that are dominated by continuum, with few absorption features. In this paper, all phases are given in rest-frame days relative to peak brightness in our synthetic 400\,nm filter.

We further select our SLSN~I sample to only contain events with a single main peak in the light curve, as otherwise there can be ambiguity in identifying the main peak and measuring phases. 
We therefore exclude those events with a \lq secondary peak\rq\/ \citep[e.g. iPTF13dcc, iPTF15esb;][around 5\% of the literature SLSNe~I]{vr17,yan17b}, where a \lq secondary peak\rq\/ is defined as a peak, before or after the brightest (main) peak, showing an absolute magnitude difference of $\lesssim 1$ mag with respect to the main peak. We also remove events with H$\alpha$ in their spectra \citep[e.g., iPTF13ehe;][]{yan15}, another 5\% of SLSNe~I in the literature. Note we do not exclude {\it a priori} SLSNe~I with early-time \lq bumps\rq\/ (see Table~\ref{table:datano}).

Around 50\% of the published SLSN~I events in the literature pass our requirements, and these can be found in Table~\ref{table:data}, while those rejected are reported in Table~\ref{table:datano} together with the criterion by which they are removed from the sample.

\begin{figure}
\centering
\includegraphics[width=\columnwidth]{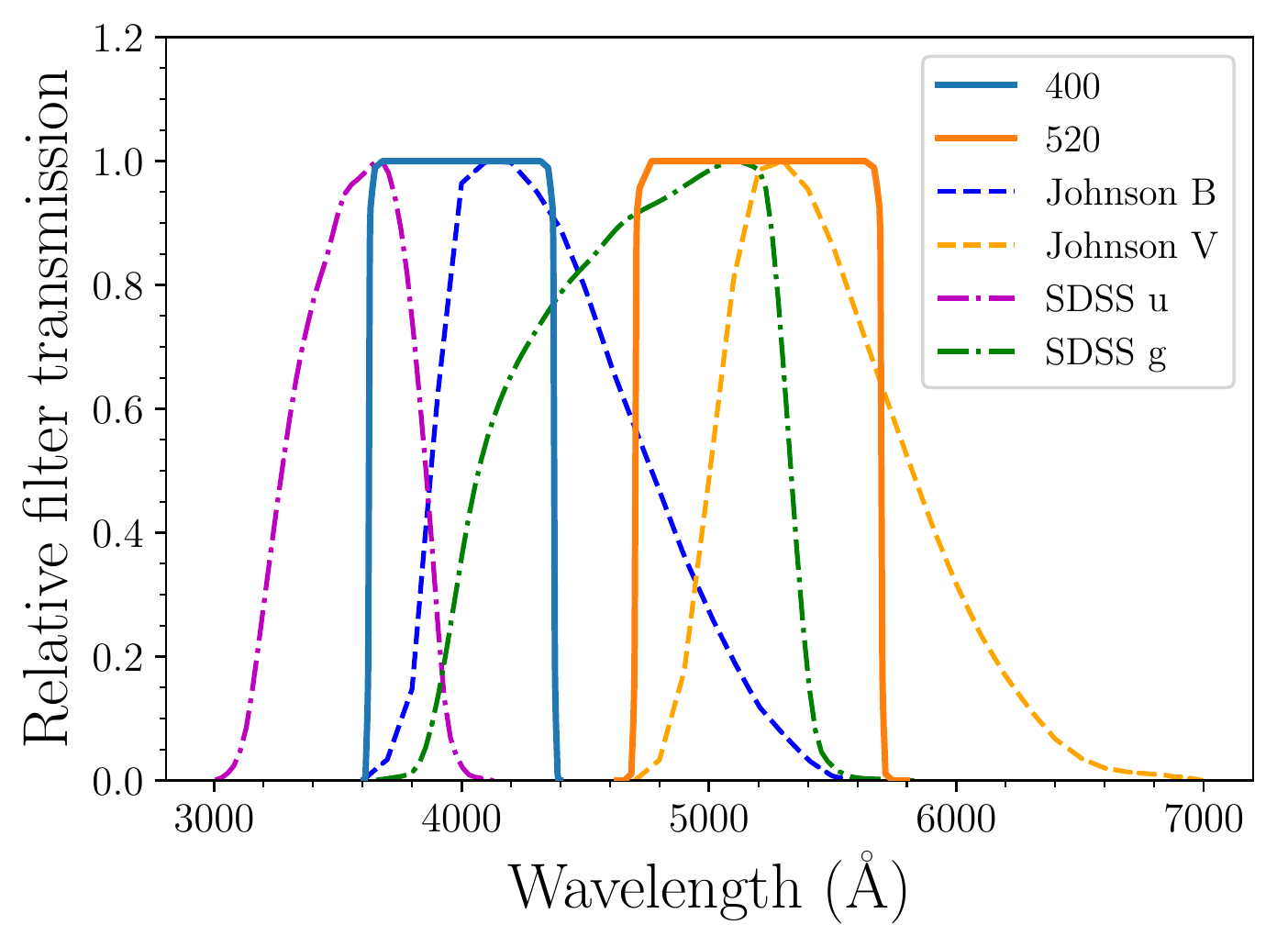}
\caption{The synthetic box filters at 4000\,\AA\ and 5200\,\AA\ (solid lines), together with the closest Johnson bands ($B$ and $V$ respectively; dashed lines), and the closest SDSS filters ($u$ and $g$; dash-dot lines). The box filters have widths of 800\,\AA\ and 1000\,\AA\ respectively.}
\label{fig:bands}
\end{figure}

We $k$-correct all published photometry for each object to our two synthetic filters (400\,nm and 520\,nm), calculated with the \textsc{snake}\footnote{https://github.com/cinserra/S3} software package \citep{in16}, which also estimates the uncertainties on the $k$-corrections. The synthetic filters, together with the standard Bessell $B$/$V$ filters and Sloan Digital Sky Survey (SDSS) $u$/$g$ filters, are shown in Figure~\ref{fig:bands}. 
Applying $k$-corrections from arbitrary observed filters over $0.1<z<4.0$ to $B$ and $V$ instead of the two box filters would result in a difference of $-0.03$\,mag at peak epoch (both $400-B$ and $520-V$), and $0.01$\,mag ($400-B$) and $0.05$\,mag ($520-V$) around 30 days after rest-frame maximum. When observed spectra for a specific SLSN are not available, we use an average SLSN~I time-series spectral energy distribution (SED), based on the methodology of \citet{pr17}. We correct all our observed photometry for Milky Way extinction prior to $k$-correction using the prescriptions of \citet{sf11}, but make no corrections for extinction in the SN host galaxies, which is believed to be small \citep[e.g.][]{ni15,le15}. Finally, we convert the rest-frame apparent magnitudes into absolute magnitudes using a flat $\Lambda$CDM cosmology, with $H_{0}=72$\,km\,s$^{-1}$\,Mpc$^{-1}$, $\Omega_\mathrm{matter}=0.27$, and $\Omega_{\Lambda}=0.73$. 

\begin{deluxetable*}{l|c}
%\tablewidth{0pt}
\tablecaption{SLSNe~I that did not pass our selection criteria.\label{table:datano}}
\tablehead{
\colhead{Selection criterion} & \colhead{SLSNe~I}}
\startdata
 Light-curve sampling & SN2006oz (1), SN2007bi (2),  SNLS06D4eu (3), SNLS07D2bv (3), PTF09cwl (4), PTF09atu (4),\\ 
 & PTF10hgi (5), PS1-10awh (6), DES14X3taz (7), DES15E2mlf (8), SSS120810:231802-560926 (9),\\
 & LSQ14bdq (10), LSQ14an (11), SN2017egm (12) \\
 Double peak & iPTF13dcc (13), iPTF15esb (14)\\
 Late time \Ha & iPTF13ehe$^\dagger$ (15), iPTF16bad (14)\\
 Chauvenet's criterion & DES13S2cmm (16), PS1-14bj (17)\\
\enddata
 \tablecomments{References: 1. - \citet{le12} - 2. \citet{gy09} - 3. \citet{ho13} - 4. \citet{qu11} - 5. \citet{in13} - 6. \citet{cho11} - 7. \citet{smi16} - 8. \citet{pan17} - 9. \citet{ni14} - 10. \citet{ni15a} - 11. \citet{in17} - 12. \citet{bose17} - 13. \citet{vr17} - 14. \citet{yan17} - 15. \citet{yan15} - 16. \citet{papa15} - 17. \citet{lu16} \\
 %$\dagger$ Used in the peak luminosity color test\\
 $\dagger$ Used as a test in the four observables parameter space (4OPS)}
\end{deluxetable*}

\begin{figure}
\centering
\includegraphics[width=\columnwidth]{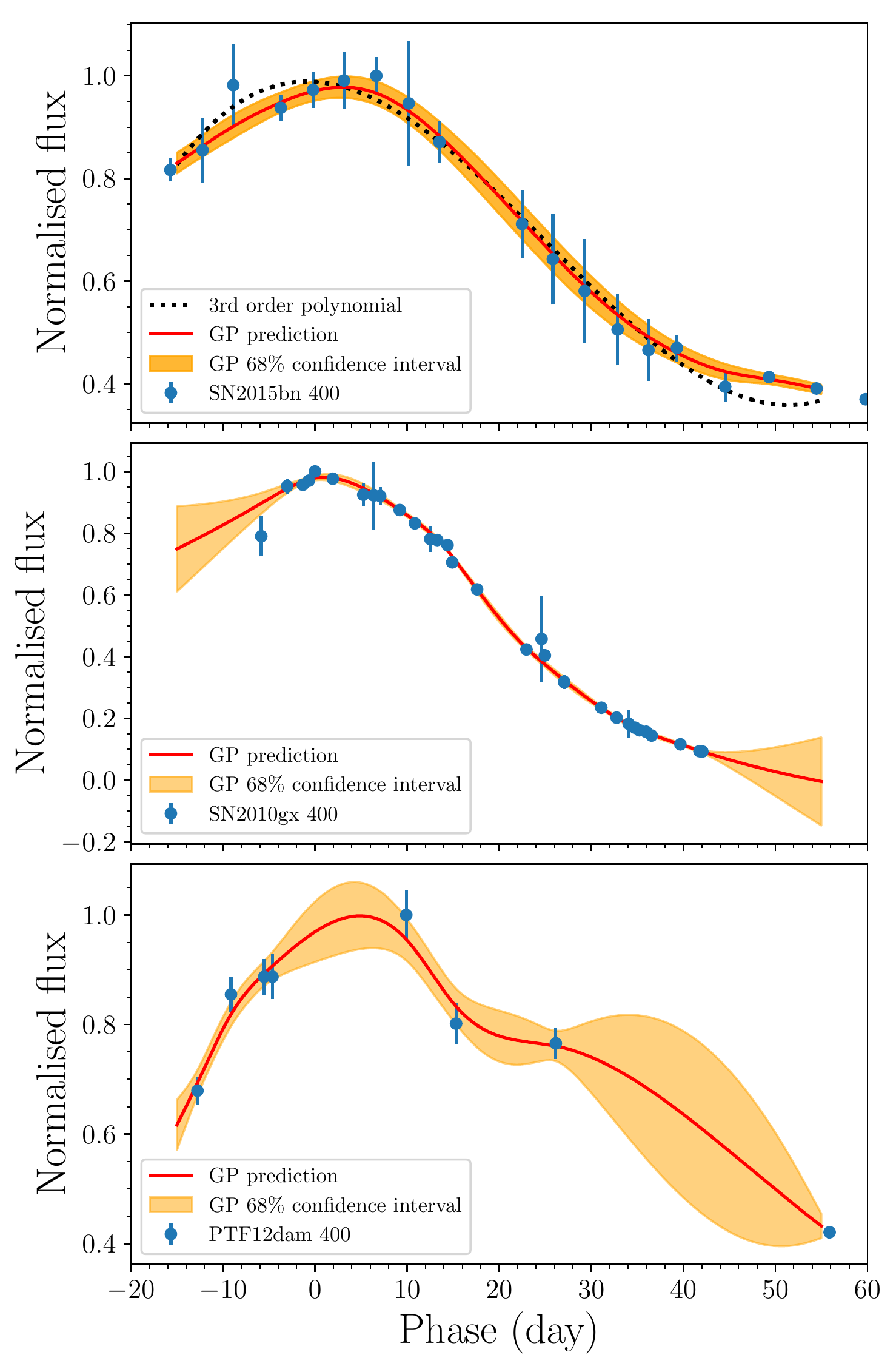}
\caption{Upper panel: Gaussian process (GP) fitting of SN2015bn, with the \textsc{george} machine learning library compared to a 3rd-order polynomial (dotted line). Center and lower panels: two other example GP fits. The center panel is SN2010gx, a well-sampled event, and the lower panel is PTF12dam, a sparsely-sampled event. In all panels, the data are shown as filled circles, the GP fits are solid lines, and the uncertainty in the fit is the shaded area.}
\label{fig:GP}
\end{figure}

\subsection{Gaussian processes regression}\label{ss:gp}
\label{sec:GP}

To estimate the SLSN~I brightness around peak epoch ($-15\leq\mathrm{phase}\leq30$), where literature SLSNe~I have the most coverage, we investigate several techniques to fit the available data set, including polynomial fitting \citep[as in][]{in14}, and interpolation using Gaussian processes (GPs) regression \citep{bishop,gps}. GPs are already successfully used in several areas of astronomy \citep[e.g.][]{maha08,way09,gibson12} and in the context of supernovae \citep[e.g.][]{kim13,scalzo14,2017arXiv170901513D}. In supernova analyses, GPs can be used for Bayesian regression and mean function fitting with a non-parametric approach, e.g. broad-band light-curves, bolometric light-curve, temperature and radius evolution, as well as line profiles in spectra. 
 
A GP assumes that our variable $y$ is randomly drawn from a Gaussian distribution with a certain mean and covariance ($y\sim f(\mu,\sigma^2)$), and then considers $N$ such variables drawn from a multivariate Gaussian distribution computing their joint probability density. This is $Y\sim f(m,K)$, where $Y = (y_1, ...,y_N)^T$, $m = (\mu_1, ...,\mu_N)^T$ is the mean vector (transposed) and $K$ is the covariance matrix, called the \lq kernel\rq\/, having  a $N\times N$ dimension of the form
\begin{center}
\[
K=
  \begin{bmatrix}
    \sigma_1^2 & \dots & cov(y_1,y_N) \\
     \vdots & \ddots & \vdots \\
     cov(y_1,y_N)& \dots  & \sigma_N^2\\
  \end{bmatrix}.
\]
\end{center}
In the case of uncorrelated variables $cov(y_i,y_j)=0$, this becomes a diagonal matrix. This approach provides for a probability distribution over functions and it allows us to compute a confidence region for the underlying model of our variable. A GP is hence specified by its mean function and kernel.

To reach convergence of the distribution, the kernel hyper-parameters are optimized using the maximum likelihood method. We test several kernels to find the most suitable covariance function for the objects in our SLSN sample. The kernels we consider are an exponential sine squared kernel (suited for periodic functions), a linear kernel, a polynomial kernel and a squared exponential kernel. As a basic metric for quality of fit we compute the $\chi^2$, comparing the mean of the GP posterior distribution to our data. We find that the \lq Matern-3/2\rq\ kernel gives the best fit. The kernel can be written in terms of radius, $r = | a_i - a_j |$, as:
\begin{equation}
f(r)= c^{2} \, \left( 1 + \frac{\sqrt{3}\,r}{t} \right) exp\left( - \frac{\sqrt{3}\,r}{t} \right),
\end{equation} 
where $c$ and $t$ are the best fit hyper-parameters.

We use the Python package {\sc george} \citep{george} to perform our GP regression. Figure~\ref{fig:GP} (top) shows a comparison between the Matern-3/2 kernel and a third-order polynomial fit. The GP fit is a better representation of the data with respect to the polynomial, meaning a lower $\chi^2$.
Figure~\ref{fig:GP} also shows examples of the GP fits to 400\,nm data, while all the fits are reported in Appendix~\ref{sec:GPfits}. As expected, when the data are sparse the fit uncertainties are larger (bottom panel). A key advantage of the GP fitting is that the fit uncertainties, as a function of phase, are naturally produced by the GP fitting. Accurate uncertainties will be important for the analysis in this paper. We further refer to \citet{sdm2014} for a more in-depth analysis of advantages and drawbacks of GPs in astronomy. 

We fit our entire SLSN sample 
from -15\,d (in the 400\,nm band) to +55\,d. Whenever the data or GP uncertainties on the magnitude are smaller than those reported by the survey that discovered the SLSN, we replace them with the typical survey photometric uncertainties at the redshift of the SN (e.g., for the 400\,nm band, PS1 averaged uncertainties at $z <0.25$ and $0.25<z<0.60$ are 0.02 and 0.06\,mag, while the DES uncertainties at $z<0.60$ and $0.60<z<0.90$ are 0.04 and 0.05\,mag)\footnote{Derived from SLSN PS1 and DES papers listed in Tables~\ref{table:data}~\&~\ref{table:datano} and from DES private communications.}. The resulting fit magnitudes are reported in Table~\ref{table:data}, together with their uncertainties.

\begin{deluxetable*}{lcccccccccc}
\tabletypesize{\scriptsize}
\tablewidth{0pt}
\tablecaption{Sample of SLSNe~I. Associated errors in parentheses.\label{table:data}}
\tablehead{
\colhead{SN} & \colhead{$z$} &\colhead{Ref.$^a$} & \colhead{Type } & \colhead{$M$(400)$_{\rm 0}$} & \colhead{$M$(400)$_{\rm 0}$-$M$(520)$_{\rm 0}$} 
& \colhead{$\Delta M_{20}$(400)} & \colhead{$\Delta M_{30}$(400)} & \colhead{$M$(400)$_{\rm 30}$-$M$(520)$_{\rm 30}$} & \colhead{$v_{\rm 10}$\,\,$^b$} & \colhead{ $\dot{v}$\,\,$^c$}
}
\startdata
Gaia16apd &0.102&1&fast & -21.87 (0.04)&-0.18 (0.07)	&0.69 (0.06)&1.30 (0.08)&0.28 (0.07)&13200 (2000)&50 (70)\\	
PTF12dam&0.107&2& slow	 &-21.70 (0.07)	&-0.23 (0.06)	&0.31 (0.09)&0.40 (0.18)&-0.09 (0.12)&9500 (1000)&5 (50)\\	
SN2015bn &0.114&3& slow	 &-21.92 (0.02)	&-0.15 (0.04)	&0.36 (0.05)&0.66 (0.06)&0.04 (0.07)&9000 (1000)&25 (45)\\
SN2011ke&	0.143&2&fast	 & -21.23 (0.09)	&	0.04 (0.13)	&0.89 (0.09)&1.63 (0.09)&0.59 (0.03)&17800 (2000)&280 (75)\\
SN2012il	&      0.175&2&fast &	-21.54 (0.10)&	-0.02 (0.11)	&1.39 (0.17)&1.65 (0.17)&0.48 (0.13)&17500 (2000)&242.5 (100)\\
PTF11rks	&      0.190&2&fast	& -20.61 (0.05)	&0.20 (0.06)&0.87 (0.07)&2.11 (0.11)&1.16 (0.15)&17200 (2000)&110 (100)\\
SN2010gx&	0.230&2&fast &	-21.73 (0.02)&	-0.11 (0.02)	&0.76 (0.03)&1.55 (0.04)&0.53 (0.03)&18500 (2000)&260 (100)\\
SN2011kf&	0.245&2&fast	 &-21.74 (0.15)	&-	&0.52 (0.18)&1.03 (0.21)&-&-&-\\
LSQ12dlf&	0.255&2&	fast &-21.52 (0.03)	&0.05 (0.03)&0.76 (0.04)&1.27 (0.11)&0.57 (0.10)&15600 (1000)&145 (32.5)\\
LSQ14mo &0.256&4,5&fast	 &-21.04 (0.05)	&-0.08 (0.04)	&1.30 (0.14)&2.23 (0.14)&0.61 (0.02)&14000 (1800)&130 (82.5)\\
PTF09cnd&	0.258&2&fast	 &-22.16 (0.08)	&-	&0.71 (0.14)&1.04 (0.12)&-&-&-\\
SN2013dg&	0.265&2&fast	 &-21.35 (0.05)	&-0.26 (0.08)&1.03 (0.06)&1.90 (0.08)&0.56 (0.10)&15700 (1000)&265 (55)\\
SN2005ap&	0.283&2&fast	 &-21.90 (0.04)	&-	&0.85 (0.09)&-&-&-&-\\
PS1-11ap &0.524&2& slow	 &-21.78 (0.03)	&-0.25 (0.03)	&0.35 (0.04)&0.67 (0.05)&-0.06 (0.05)&8800 (2500)&15 (87.5)\\	
PS1-10bzj&	0.650&2&fast	 &-21.03 (0.06)	&0.15 (0.11)&1.23 (0.32)&1.82 (0.26)&0.94 (0.25)&-&-\\
iPTF13ajg &0.740&6&fast	 &-22.42 (0.07)	&-0.29 (0.09)	&0.19 (0.10)&0.45 (0.10)&-0.11 (0.09)&15500 (2000)&100 (100)\\
PS1-10ky	&      0.956&2&fast	 &-22.05 (0.06)	&-0.06 (0.07)&0.61 (0.07)&1.20 (0.07)&0.25 (0.06)&-&-\\
SCP-06F6&	1.189&2&fast	 &-22.19 (0.03)	&-	&0.57 (0.15)&0.96 (0.30)&-&-&-\\
PS1-11bam & 1.565 & 7,8&  -	 &-22.45 (0.10)	& -	& 0.36 (0.14)&0.60 (0.14)&-&-&-\\
\cutinhead{Test}
iPTF13ehe &0.343&9&slow	 &-21.58 (0.04)	&	-0.29 (0.05)	&0.08 (0.06)&0.22 (0.06)&-0.10 (0.05)&10600 (2300)&-\\
\cutinhead{Outliers}
PS1-14bj &0.521&10& - &-20.44 (0.05)	&	0.22 (0.06)	&0.03 (0.07)&0.07 (0.07)&-0.01 (0.07)&-&-\\	
DES13S2cmm &0.663&11& - &-20.41 (0.05)	&	-0.24 (0.06)	&0.83 (0.09)&0.96 (0.13)&0.23 (0.14)&-&-\\	
 \enddata
 \tablecomments{$^a$ References for observed light curves and spectra: 1. \citet{kan17} -- 2. \citet{in14} and references therein -- 3. \citet{ni16} -- 4. \citet{ch17a} -- 5.\citet{le15b}  -- 6. \citet{vr14} 
 -- 7. \citet{berger12} -- 8. \citet{lu17}
 -- 9. \citet{yan15} -- 10. \citet{lu16} -- 11. \citet{papa15} .\\
 $^b$ \kms, measured from \ion{Fe}{2} $\lambda$5169.\\
 $^c$ \kms\/ day$^{-1}$, $\Delta v/\Delta t$, where $\Delta t$ is measured from +10 to +30\,d using \ion{Fe}{2} $\lambda$5169.}
\end{deluxetable*}

%%%%%%%%%%%%%%%%%%

\subsection{Line velocity measurements}\label{ss:lv}
\label{sec:velocities}

Our final measurements concern the SLSN spectra, and in particular the estimation of the photospheric velocities. In core collapse SNe, \ion{Sc}{2} $\lambda$6246, and subsequently \ion{Fe}{2} $\lambda$5169, are the best available proxies to trace the photospheric evolution due to their small optical depth \citep{branch02}. In SLSN~I spectra, \ion{Sc}{2} is not visible, but \ion{Fe}{2} has been measured using several different approaches \citep[e.g.][]{in13,ni15,liu17}.

We measure the line velocities in all spectra from +10 to +30\,d. Before +10\,d, the ionic component is weak \citep{in13} and contaminated by \ion{Fe}{3} \citep{liu17}. When spectra around $10\pm2$\,d and $30\pm2$\,d are not available for a given SLSN, we estimate the velocity from a least-squares fit of the measurements from nearby epochs and account for an additional uncertainty in the estimate using $\sigma_{\rm final} = \Delta t / \sigma_{\rm measure}$, 
where $\Delta t$ is the phase difference (in seconds) between the estimated and measured epochs. Only 12 out of the 19 SLSNe have spectra covering the wavelength region and the time-frame of interest (see Table~\ref{table:data}). 

We measure the velocity from fits to the absorption minima. We experiment with three different profiles for the fits (Gaussian, skewed Gaussian, and Voight), finding an overall agreement among the three profiles. We repeat the measurements several times for each feature, changing the continuum levels to better estimate the uncertainties. We then use the mean of the measurements as the final value, and the standard deviations as the uncertainty estimate; the values are tabulated in Table~\ref{table:data}. This approach has been widely used in measuring the line velocities of SLSNe~I, with consistent results \citep[e.g.][]{pasto10,cho11,in13,smi17}.

We have also cross-checked our velocity measurements using a GP approach (Section~\ref{sec:GP}), fitting the wavelength region from 4800\,\AA\ to 5200\,\AA, and finding the local minimum. We use the same Matern-3/2 kernel as with the light curve fitting. As would be expected, given the greater flexibility, the GP fits usually return a $\chi^2$ comparable to or lower than the standard fitting procedure, with the advantage of improving fitting/deblending multi-component profiles without a biased prior knowledge of the feature types and numbers (e.g., absorptions/emissions with P-Cygni profiles, Lorentzian or Voight wings, etc.). However, to properly evaluate the uncertainties we need a kernel function based on the uncertainties in the flux of the spectra. This information is missing for $\sim$40\% of our data set, and hence we use the profile fitting to allow for consistency in the approach.

Our measurements and line evolutions are broadly in agreement with those of \citet{liu17}, those reported in the papers listed in Table~\ref{table:data}, and the photospheric velocities reported by the modeling of \citet{maz16}. The only noticeable difference is in the velocity of Gaia16apd, where we found a decrease of $\sim$1000\,\kms\ over the phase range analyzed. This is due to the presence of galaxy lines that make the fit more complicated and biased by the choice of the number of components to analyze. 
The \ion{Fe}{2} $\lambda$5169\,\AA\ velocity measurement is reported as that at +10\,d, or $v_\mathrm{10}$ (\kms), while the velocity evolution over the phase range from 10\,d to 30\,d post-peak is  $\dot{v}= -\Delta v/ \Delta t$ (\kms\,day$^{-1}$), in a similar fashion to that used in SNe Ia \citep{be05}.

\section{The Four Observables Parameter Space (4OPS)}\label{sec:4ops}

\begin{figure*}
\center
\includegraphics[width=0.9\textwidth]{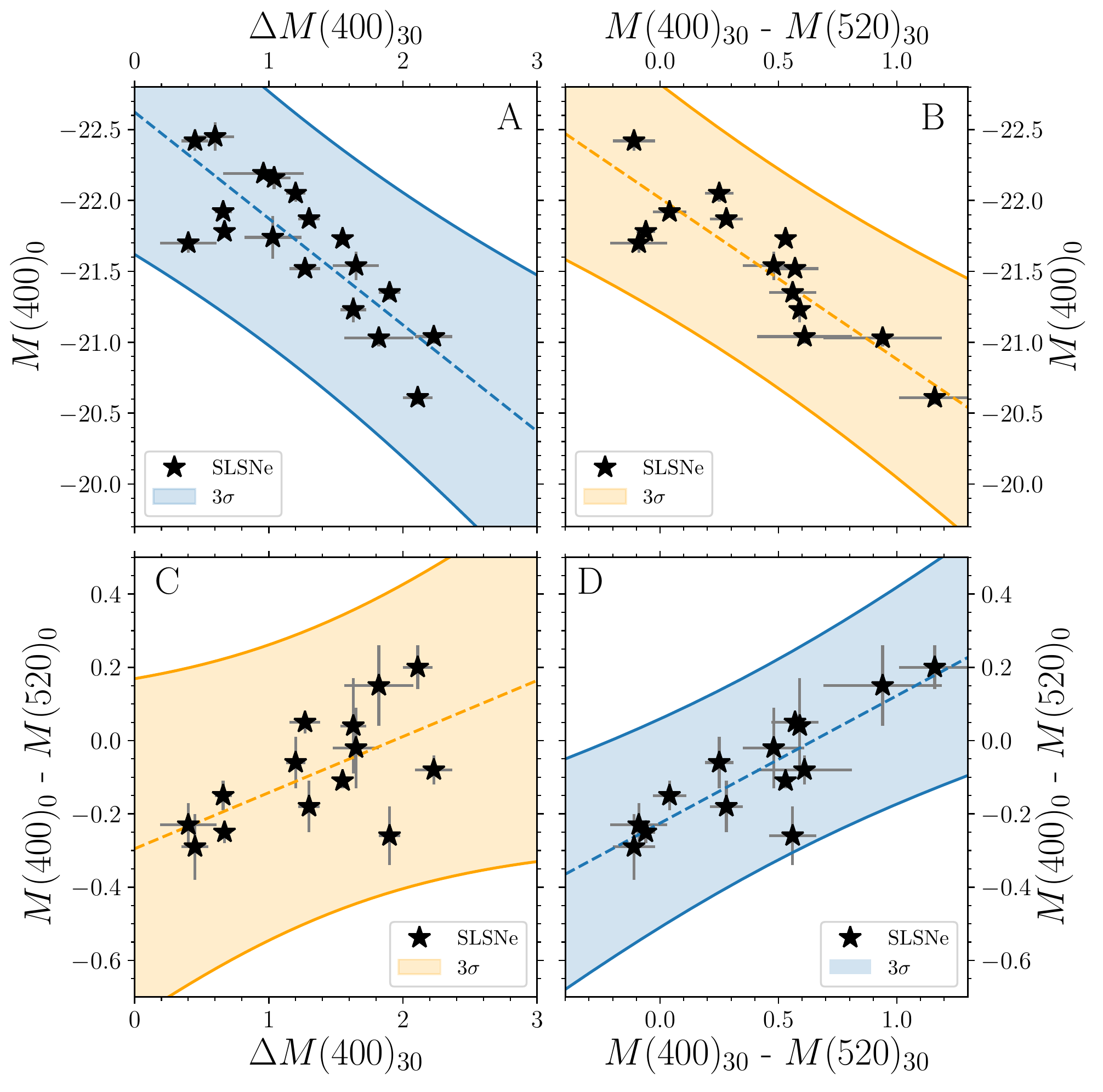}
\caption{The Four Observables Parameter Space (4OPS) plot. Top left: peak luminosity of our literature SLSN~I sample in the 400\,nm band (\bluemag{0}) versus the decline in magnitude over 30 days $\Delta\bluemag{30}$. Top right: \bluemag{0} versus color at $+30$\,d (\thirtycol). Bottom left: peak color (\peakcol) versus $\Delta\bluemag{30}$. Bottom right: \peakcol\ versus \thirtycol. 99.72\% confidence bands from the Bayesian linear regression are also shown for each panel. The four plots allow the definition of a main population of SLSN I regardless of the peak luminosity. A SLSN~I will belong to the main population if it falls in the confidence interval for the blue areas in the A and D panels or, alternatively, in the orange areas of the B and C panels. 4OPS can also be used to predict missing observables for a SLSN.}
\label{fig:4OPS}
\end{figure*}

\begin{deluxetable*}{ccccccccc}
\tablewidth{0pt}
\tablecaption{Fit parameters and statistical results of the four observables parameter space relations\label{table:fit}.}
\tablehead{4OPS Panel &$x$ & $y$ &\colhead{N (objects)}& \colhead{$\beta$} & \colhead{$\alpha$} & \colhead{$\sigma$} & \colhead{Variance} & \colhead{Pearson}}
\startdata
A & $\Delta M$(400)$_{\rm 30}$ & $M$(400)$_{\rm 0}$ & 18 & $-22.62\pm0.21$& $0.75\pm0.15$ & $0.32\pm0.23$ & $0.10\pm0.05$ & $0.82\pm0.11$ \\
B & $M$(400)$_{\rm 30}$ - $M$(520)$_{\rm 30}$ & $M$(400)$_{\rm 0}$ & 14 & $-22.02\pm0.13$& $1.14\pm0.26$ & $0.29\pm0.24$ & $0.08\pm0.05$ & $0.87\pm0.11$ \\
C & $\Delta M$(400)$_{\rm 30}$ & $M$(400)$_{\rm 0}$ - $M$(520)$_{\rm 0}$ & 14 & $-0.30\pm0.11$& $0.16\pm0.07$ & $0.14\pm0.12$ & $0.02\pm0.01$ & $0.62\pm0.24$ \\
D & $M$(400)$_{\rm 30}$ - $M$(520)$_{\rm 30}$ & $M$(400)$_{\rm 0}$ - $M$(520)$_{\rm 0}$ & 14 & $-0.22\pm0.04$& $0.03\pm0.09$ & $0.08\pm0.08$ & $0.01\pm0.01$ & $0.84\pm0.13$ \\
\enddata
\tablecomments{Least squares fits for a Bayesian weighted linear regression with weighted errors both in $x$ and $y$ of the form $\eta$ = $\beta$ + $\alpha\times x'$ + $\epsilon$, where  $x = x' + x_{\rm err}$ and  $y = \eta + y_{\rm err}$. The $\sigma$ is the standard deviation (in $y$) of this fit. The last column gives the Pearson correlation coefficient $r$.}
\end{deluxetable*}

Having assembled our SLSN~I data sample in Section~\ref{sec:sample}, we now investigate methods for classifying the events based mainly on photometric data. Our light curve fitting has provided smooth time-dependent light curves in the 400\,nm and 520\,nm filters, together with realistic estimates of the uncertainties at interpolated epochs. We select various observational quantities for which we can explore the classification potential, based on the inferred decline rates (cf. the $\Delta M_{15}$ quantities used in studies of SNe Ia), peak magnitudes, and colors. Specifically, we use
\begin{enumerate}
\item The peak luminosity in the 400\,nm filter, \bluemag{0};
\item The decline in magnitudes in the 400\,nm filter over the 30 days following peak brightness, $\Delta\bluemag{30}$; 
\item The $400-520$ color at peak, \peakcol;
\item The $400-520$ color at $+30$ days, \thirtycol.  
\end{enumerate}
These four observational quantities are tabulated in Table~\ref{table:data}, and we visualize the relationships between them in Figure~\ref{fig:4OPS}, which we term the \lq four observables parameter space\rq\ (or 4OPS).

\begin{figure*}
\center
\includegraphics[width=0.31\textwidth]{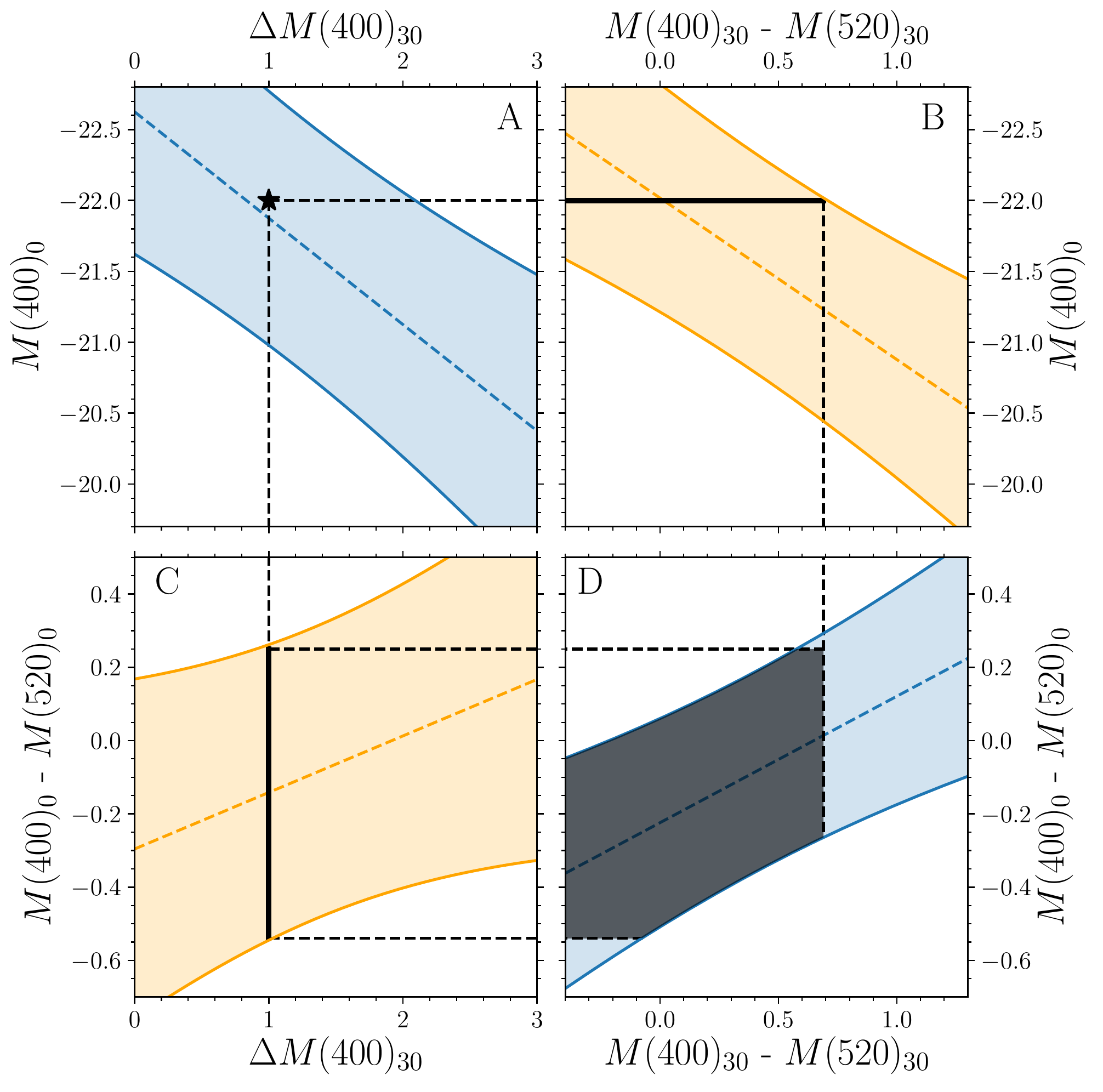}
\includegraphics[width=0.31\textwidth]{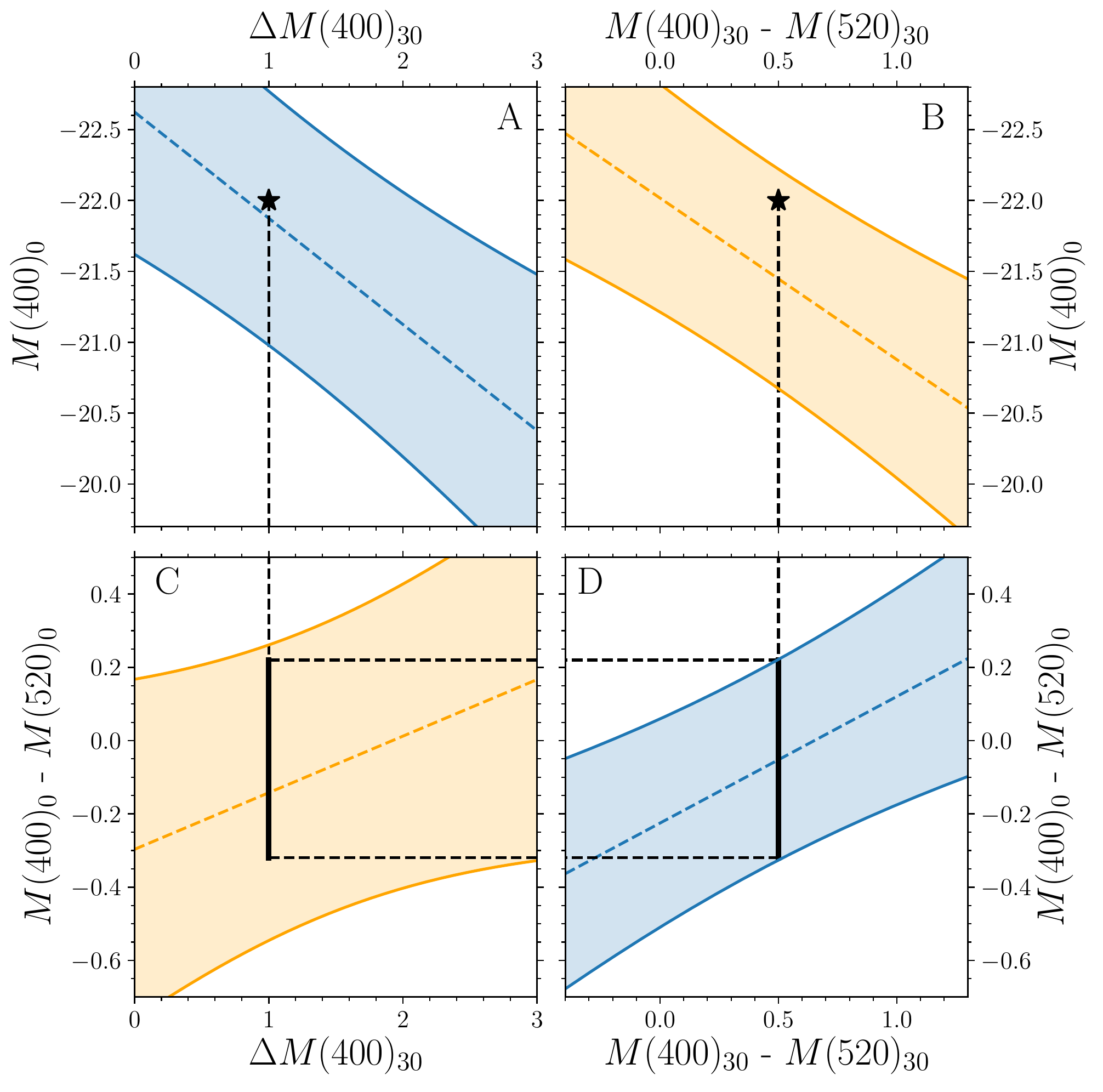}
\includegraphics[width=0.31\textwidth]{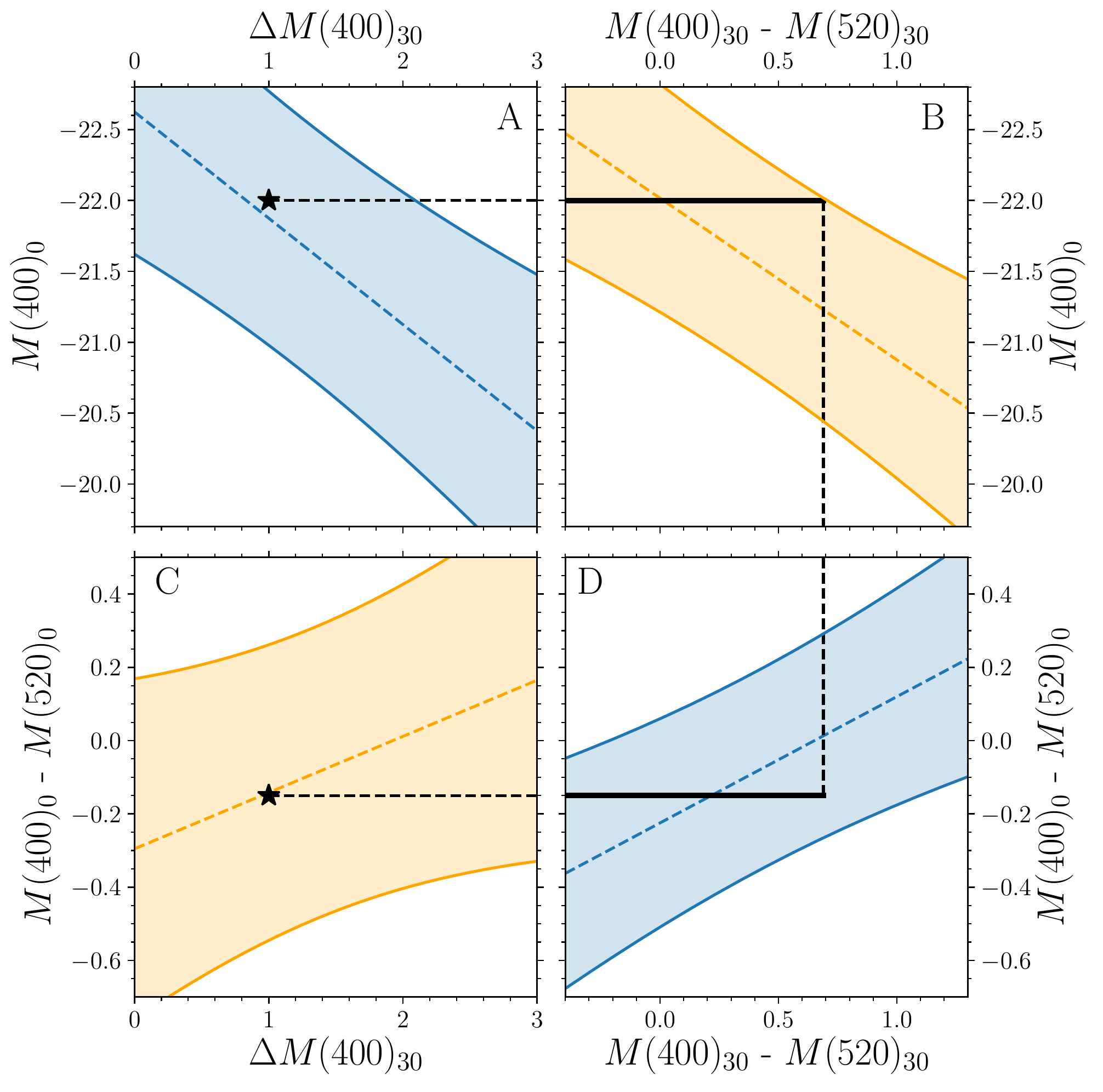}
\caption{The Four Observables Parameter Space (4OPS) plot predictions. Left: if information is available only for panel A, the prediction in panel D is the black shaded area of. Middle: if information for panels A and B is available, the predictions in panels C and D is shown (solid black line). Right: for panels A and C, the prediction in B and D is shown (solid black line).}
\label{fig:4OPSprediction}
\end{figure*}

We use a Bayesian approach to evaluate a linear regression of these parameters, allowing for the uncertainties in both the $x$ and $y$ variables (see Section~\ref{ss:gp}) and any intrinsic scatter \citep[see][for further details]{ke07}. This process uses Bayesian inference that returns random draws from the posterior. Convergence to the posterior is performed using a Markov chain Monte Carlo with $10^5$ iterations. We note that the probability of retrieving a slope $\alpha=0$ from the random draws in our fits is 0\%, i.e., the correlations are highly significant.  

As a final quality check before the definition of a likelihood area we use Chauvenet's criterion, which is a statistical procedure that provides an objective and quantitative method for data rejection based on the standard deviation of a distribution. It compares the absolute value of the difference between the suspected outliers and the mean of the sample divided by the sample standard deviation. 
We apply that in the light-curve and color evolution space, and identify two such outliers,  DES13S2cmm and PS1-14bj, which for example have a $\delta y_{\rm (theory - measure)} / \sigma$ of  $\sim$5 (DES13S2cmm) and $\sim$7 (PS1-14bj) for the peak-decline relation (see Panel A of Figure~\ref{fig:4OPS}), and hence greater than the Chauvenet threshold of 2.20, valid for a sample of 18 objects (cf. Table~\ref{table:data}).

Using the weighted linear regression fits on our final sample (see Table~\ref{table:data}) and their standard deviation ($\sigma$), we define a 3$\sigma$ region as the likelihood area in which our sample of SLSNe~I lie (see Table~\ref{table:fit} for the fit parameters). We use this area to define the photometric properties of a SLSN~I events -- by construction, it includes all SLSNe~I in our sample with sufficient photometric sampling and that do not exhibit peculiarities such as clear interaction or double-peaked light curves.

In Figure~\ref{fig:4OPS}, the two sets of diagonal panels (i.e., panels A/D and panels B/C) each display information from all four variables, and thus contain complementary information; the adjacent panels (both horizontally and vertically) contain ancillary information. The adjacent panels can also be used to predict the values (with a 3$\sigma$ uncertainty) of the other two missing variables. For example, if the peak luminosity (\bluemag{0}) and luminosity decline ($\Delta\bluemag{30}$) are measured, the SLSN colors at peak and at $+30$ days can be reliably estimated (left of Figure~\ref{fig:4OPSprediction}). If we have information on three out of four observables, we can predict the fourth one with a higher precision, namely 3$\sigma$ of the strongest among the two correlations using the missing observables (middle and right in Figure~\ref{fig:4OPSprediction}). This could be useful in current and future surveys when a band, or measurement, is not sampled due to redshift, cadence or adverse weather.

A SLSN~I belonging to the main population has to be in both the blue regions in Figure~\ref{fig:4OPS} (A and D panels) or, alternatively, in both the two orange regions perpendiculars to the blue (B and C panels). This allows us to define the bulk
of the SLSNe~I without any arbitrary magnitude limit. As a consequence, the hypersurfaces can be used to identify/classify objects as SLSNe~I in future and current surveys \citep[e.g. PS1, PTF and DES SN sample][Angus et al. in prep.]{lu17,2017arXiv170801623D} when a spectroscopic evolution is not available. However, other peculiar objects can populate the same parameter space (see next two paragraphs) and hence at least a spectrum might be warranted (see Section~\ref{sec:vel}).

We note that the outliers represent 5\% of the full literature SLSN~I sample and 9\% of those passing the selection criteria in Section~\ref{ss:sample}. This implies that the definition of a SLSN~I is achieved with a confidence level of at least 90\% which, according to Dixon's Q test, is statistically significant.

\begin{figure}
\center
\includegraphics[width=\columnwidth]{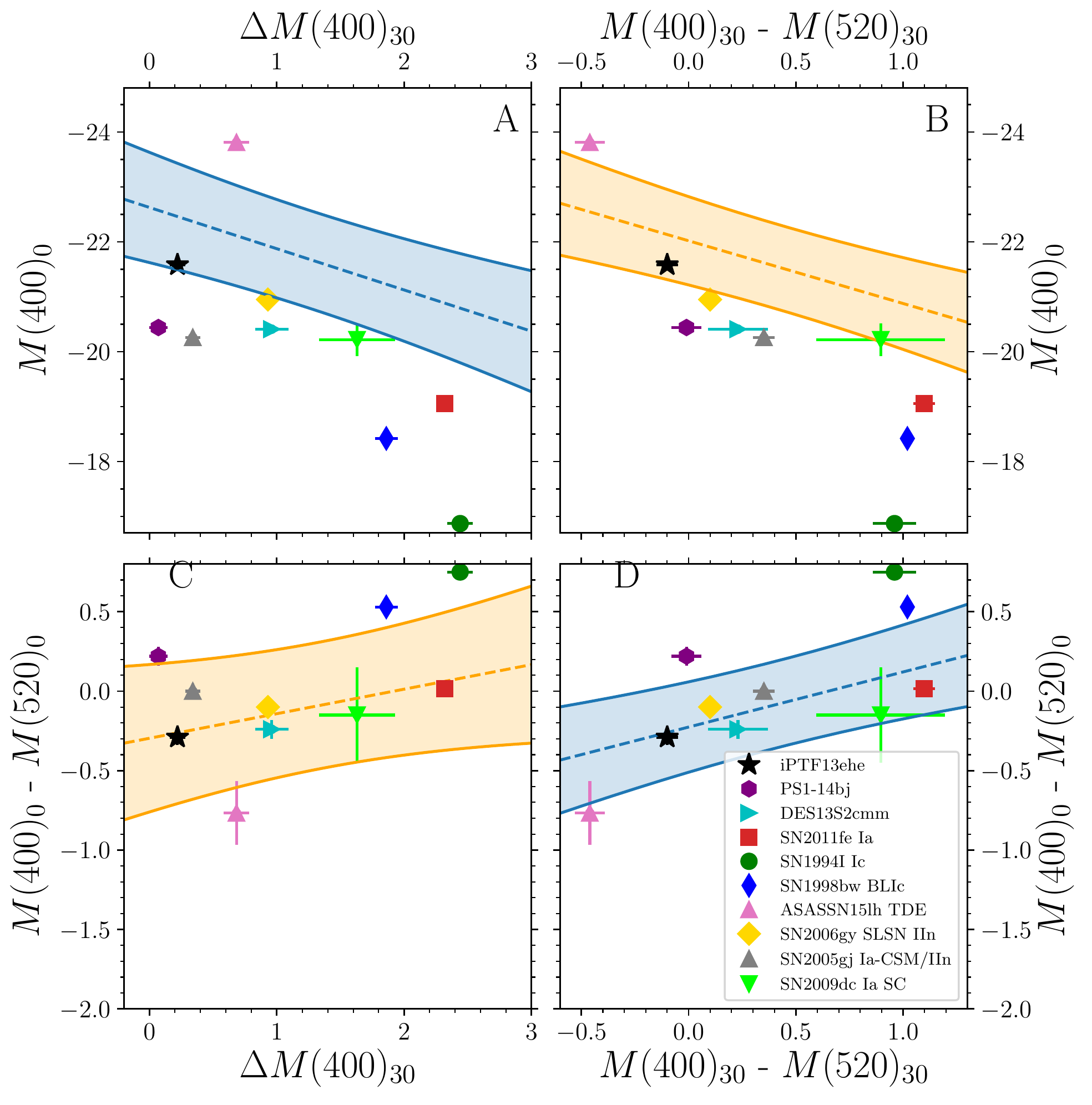}
\caption{Our SLSNe I test object (iPTF13ehe), the outliers PS1-14bj and DES13S2cmm, and various other SN types in the parameter space of Figure~\ref{fig:4OPS}. The only type of SN that could appear in the same region as SLSNe~I are the very bright type IIn (SLSNe~IIn) and possibly superchandra type Ia (Ia SC). Data references: type Ia SN2011fe \citep{brown14,pereira13}; type Ic SN1994I \citep{richmond96}; type broad-line (BL) Ic SN1998bw \citep{pat01}; tidal disruption event (TDE) ASASSN-15lh \citep{dong16,le16}; hydrogen-rich interacting SLSN IIn S2006gy \citep{smi07}, type Ia-CSM/IIn SN2005gj \citep{aldering06,prieto07}, superchandra (SC) type Ia SN2009dc \citep{tauben11}}.
\label{fig:4OPStest}
\end{figure}

To test this approach, we measured the same quantities for a literature SLSN~I showing an \Ha\ profile and slower light curve after 150 days, namely iPTF13ehe \citep[][]{yan15}. 
We find iPTF13ehe lies in all the areas and close to three slow-evolving SLSNe~I (see Figure~\ref{fig:4OPStest} and Table~\ref{table:data}). In this case we infer that iPTF13ehe, despite its late-time behavior, is consistent with a main population SLSN I. 

Figure~\ref{fig:4OPStest} also shows a SLSN~I outlier (PS1-14bj) together with other literature SNe of all types, corrected for Galactic and host extinction. Only SLSNe IIn, such as SN2006gy, and potentially some super-Chandrasekhar (SC) type Ia SNe (e.g., SN2009dc) populate the same part of the parameter space as SLSNe I. However, the spectra of these classes appear quite different from SLSNe~I. Normal H-poor SNe, such as type Ia, Ic and broad-line Ic fall below the likelihood area in panel A, as they are fainter and do not evolve as fast as the relationship would predict. Moreover, type Ic and broad-line Ic SNe have redder colors than SLSNe~I at peak and 30 days (Panel D of Figure~\ref{fig:4OPStest}), as expected from the spectroscopic evolution of SLSNe~I that at 30 days resembles a type Ic SN at peak \citep{pasto10,in13}. Furthermore, a superluminous tidal disruption event \citep[e.g. ASASSN-15lh,][]{le16}, which has also been suggested to be a SLSN~I \citep[e.g.][]{dong16}, falls outside the likelihood area since it is brighter and bluer than the main population of SLSNe I. 

\begin{figure*}
\center
\includegraphics[width=18cm]{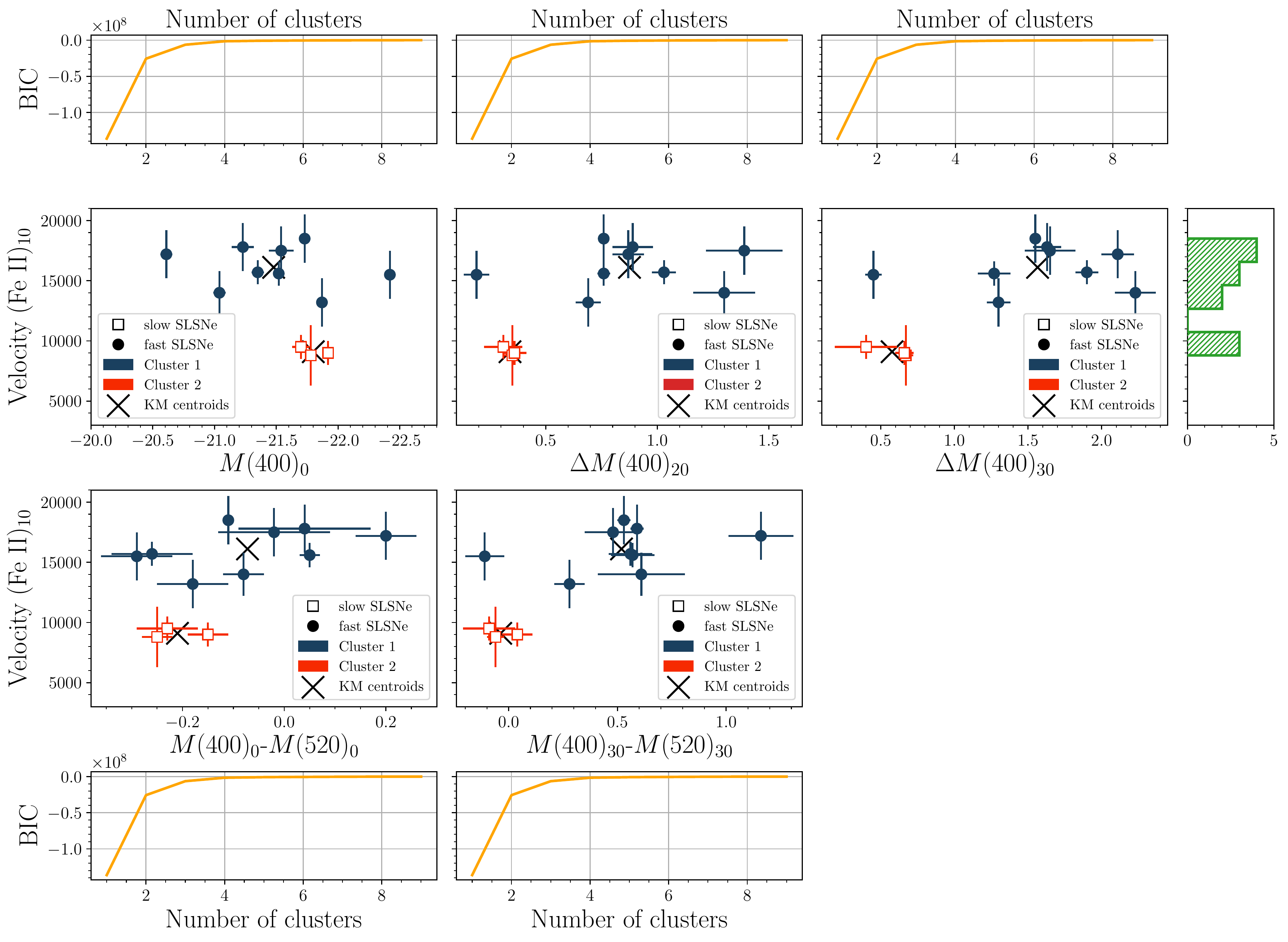}
\caption{Fe~{\sc ii} $\lambda$5169 velocities at $+10$\,d versus various photometric observables. A partitional cluster analysis using K-means methodology finds the same two classes of SLSNe~I in each plot.  BIC curves are reported for each cluster in the first and fourth row. In the fourth column we show the histogram of the velocities. The bin dimension has been chosen accordingly with Sturges' formula, which accounts only for data size and it is optimal for smaller data sets.}
\label{fig:velobs}
\end{figure*}

\section{Photospheric velocity versus photometric observables}\label{sec:vel}

As discussed in the introduction, it is unclear if both fast- and slow-evolving
SLSNe~I are two different manifestations of the same explosion mechanism, or intrinsically different transients  \citep[in terms of combination of powering mechanisms and/or progenitor scenario, e.g.][]{gy12,ni15,in17}. Combining photometric and spectroscopic measurements of a SN class can in principle reveal important physical information, or the existence of classes and/or subclasses of transients \citep[e.g.][]{hamuy03,be05,claudia17b}. To investigate we employ a similar method to that used for SNe Ia \citep{be05}, using our photospheric velocity measurements from Section~\ref{sec:velocities} (Table~\ref{table:data}).

We initially compare the \ion{Fe}{2} velocity at +10\,d with our photometric observables in Figure~\ref{fig:velobs}, using the 4OPS variables and the decline rate over 20 days in the 400\,nm filter ($\Delta\bluemag{20}$), easier to measure for high-redshift and/or fast-evolving objects.  We then perform a partitional cluster analysis, for each combination shown in Figures~\ref{fig:velobs}, using the K-means methodology.

Such a cluster analysis separates samples in groups of equal variance, minimizing the within-cluster sum of squared criterion to find the centroids of the groups \citep{kmeans}. To choose the ideal number of clusters, we initially applied a Gaussian Mixture Model using an expectation-maximization algorithm \citep{fraley2002}, and subsequently we searched for the ideal number of clusters (the K in K-means and ranging from 1 to 9) through the Bayesian Information Criterion \citep[BIC,][]{bic}, which has a probabilistic interpretation \citep{kass1995}. That creates a function ($f({\rm K})$) dependent on the number of clusters. The highest absolute value of the second derivative of the function returns the ideal number of clusters. We show the results of this test in Figure~\ref{fig:velobs}.

This statistical approach reveals the presence of two well-separated clusters in all of the spectroscopic/photometric observable parameter spaces (see Figure~\ref{fig:velobs}), allowing a natural grouping of SLSNe that can be investigated using other relationships. We also run a Monte Carlo Markov Chain with $10^5$ iterations, allowing the data to vary inside the uncertainties. We retrieve similar clusters between 95\% and 97\% of the cases, with the only exception in the peak luminosity versus the \ion{Fe}{2} velocity at +10\,d, in which we retrieved similar clusters in $\sim$90\% of the cases.

\begin{figure}
\center
\includegraphics[width=\columnwidth]{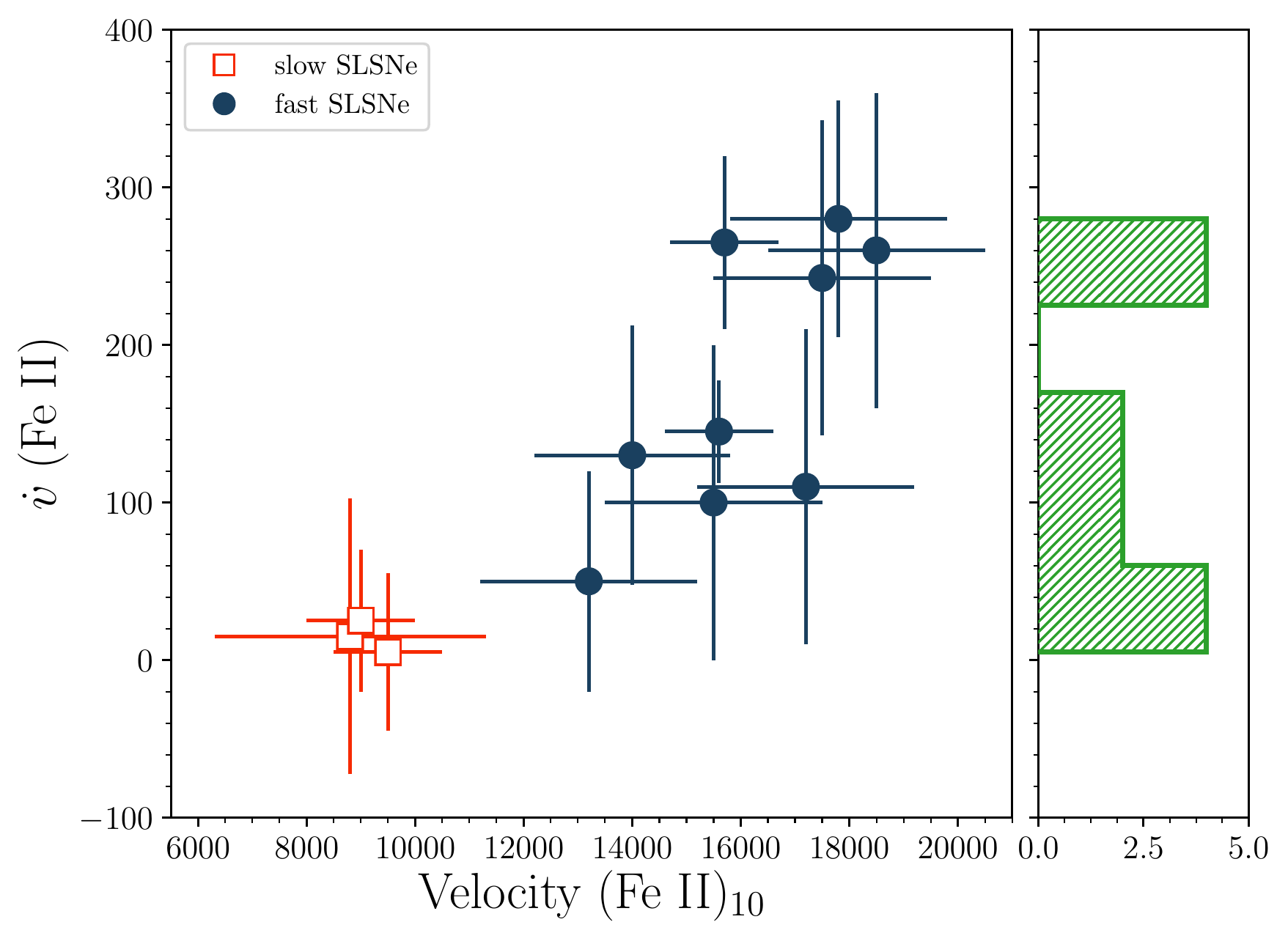}
\caption{Left panel: \ion{Fe}{2} $\lambda$5169 velocity evolution from +10 to +30 days ($\dot{v}$) versus \ion{Fe}{2} $\lambda$5169 velocities at $+10$\,d. Right panel: histogram of the velocity evolution, with the bin dimension  chosen using Sturges' formula, which accounts only for data size and it is optimal for smaller data sets.}
\label{fig:vdot}
\end{figure}

On the basis of the results of the cluster analysis, we then further investigate the spectroscopic evolution of the two clusters comparing their initial photospheric velocity with their photospheric evolution. The comparison shown in Figure~\ref{fig:vdot} suggests that the higher the photospheric velocity, the larger the gradient and hence the faster the velocity decreases. We also perform a partitional cluster analysis on the measurements of Figure~\ref{fig:vdot} finding again the same two clusters of the above analysis. Therefore, combining the information of the cluster analysis together with those of Figures~\ref{fig:velobs} and \ref{fig:vdot}, we outline the following SLSN~I subclasses:

\begin{enumerate}

\item A first group (\lq Fast\rq), consisting of SLSNe~I with fast-evolving light curves, a broad range of peak colors ($-0.3\lesssim\peakcol\lesssim0.2$), and a broad color evolution with red objects becoming redder faster (Panel D of Figure~\ref{fig:4OPS}). They have higher expansion velocities ($v_{10}\gtrsim12000$\,\kms) and large velocity gradients.

\item A second group (\lq Slow\rq), consisting of SLSNe~I with slow-evolving light curves, a narrow range of peak colors, and a color evolution of only 0.2\,mag in 30 days following peak brightness (panels in Figure~\ref{fig:4OPS}). They have lower expansion velocities compared to the fast group ($v_{10}\lesssim10000$\,\kms), and a low velocity gradient.

\end{enumerate}

We can distinguish between these two subgroups of SLSNe~I by combining almost any photometric observable with a spectrum taken around +10\,d. We applied this to iPTF13ehe, our test SLSN~I that passed the cut in the 4OPS (see Figure~\ref{fig:4OPS}) and hence defined as a main population SLSN, and found that to be clustered with the Slow group.

\section{Implications for SLSNe~I}\label{sec:dis}
\label{sec:discussion}

In this paper we have used various photometric measurements of SLSNe~I to identify a main population of SLSN events, which remains the primary purpose of our work. In this section, we discuss the implications of our results.

\subsection{Consequences of the four observables parameter space}

The parameter space of Figure~\ref{fig:4OPS} may in principle be used to help physically understand the explosion mechanisms of these transients. Relationships between the change in luminosity in one band (panel A) and the color evolution (panel B), and the broad band behavior of a SN at a given epoch (panels C \& D) are broad reflections of the physical properties of the SN ejecta (i.e. diffusion time, opacity and temperature). The correlation shown within panel A is likely a reflection of the diffusion time of the ejecta -- similar to that seen within SNe Ia \citep{phillips93}. However, as we do not consider the light curves of SLSNe~I to be radioactively driven \citep[e.g.][]{ni13,in17}, for SLSNe~I this correlation is unlikely to be solely related to the mass of the ejecta produced. 

On the other hand, the tight relation presented in panel D of the 4OPS (see Figure~\ref{fig:4OPS} and Table~\ref{table:fit}) between the color observed at peak and at +30\,d suggests that these two are correlated by one physical parameter only, which could be the temperature or the radius. 

A wide range of possibilities have been postulated to explain SLSN~I luminosities, such as the rapid spin-down of a magnetar \citep[e.g.][]{kb10,wo10,de12}, the interaction between the SN ejecta and the surrounding CSM previously ejected from the massive central star \citep[e.g.][]{ch12,woosley17}, and a pair instability explosion \citep[e.g.][]{kozyreva17}.
For all three models there are multiple parameters at play in the production of the overall luminosity and color evolution of the transient. As such, the linking of an observed behavior to a dominant physical parameter becomes complex. For example, a magnetar magnetic field strength, spin-period and explosion ejecta mass are all factors in the luminosity evolution \citep{kb10}, whilst within the interaction model, the mass of the ejecta, its density profile and distribution coupled with the mass, distance and volume of the CSM shell must be considered \citep[e.g.][]{ch13,woosley17}. 

At present there are no model predictions that aptly describe the broad-band behavior shown in Figure~\ref{fig:4OPS}. This could be due to the fact that the diffusion time not only depends upon the ejected mass, but also on the ejecta velocity and its opacity to optical-wavelength photons. Opacities, in particular, are determined by the temperature and composition of the ejecta and therefore may vary with time during the SLSN~I evolution \citep{maz16}. Thus, explaining the relation observed within any of panels A, B and C of the 4OPS with any of the above suggested scenarios is not trivial. Nonetheless, the presence of the relations hints that a pure radiative transfer in the SN ejecta should be at play and any model that aims to explain such SNe should take these observational properties into account. 

The primary purpose of our work is to define a main population of SLSNe~I. Within the context of a magnetar powered event, favored by several observational studies \citep[e.g.][]{ch15,ni15,smi16,in16a},
the behavior observed could be explained by the magnetar energy injection always occurring at a certain time (e.g., shortly after the explosion). Such a scenario would allow the diffusion time of the ejecta to be comparable to the time needed for the SN to reach peak luminosity (i.e., panel A). An injection this early would also provide the rotational energy needed to overwhelm the initial thermal energy of the SN explosion, and hence provide the energy source which drives the main peak of the light curve \citep{kb10}. Such a population of engine driven SLSNe~I would be composed of brighter objects that are overall bluer and more slowly evolving than dimmer events. Moreover, redder objects in this main population would evolve faster in both luminosity and color as inferred from Figure~\ref{fig:4OPS}, panel D, and previously shown by \citet{in14}.

Objects outside the main population of SLSNe~I, but with a luminosity evolution that could be explained by an inner engine, have already been found \citep[e.g.][ and the Dark Energy Survey collaboration, private communications]{gr15,ka16}, but their spectrophotometric evolution is different to SLSNe~I, and this is reflected in Figure~\ref{fig:4OPStest}. These objects could belong to a similar engine-driven transient family, only here the injected energy and/or the timescale over which it is injected would be somewhat different.

\subsection{Consequences of the cluster analysis and photopsheric velocity evolution}

The analysis of Section~\ref{sec:vel} returns two subclasses, which are outlined in terms of spectrophotometric evolution during the first 30 days from peak, as well as a distinct photospheric velocity behavior. In the context of the magnetar scenario, 
the almost flat velocity evolution exhibited by the slow subclass, and their overall slower velocity, suggests that the photosphere reaches the internal shell created by the magnetar bubble \citep[see eq. 7 in][]{kb10} earlier than in the fast subclass. Fast SLSNe~I with a high photospheric velocity also have a larger velocity gradient. This could be related to additional energy deposited by the magnetar into the ejecta \citep{de12}. Such energy is a function of the spin period \citep[see eq. 1 in][]{kb10}, and faster rotation would imply more energy and hence faster ejecta. 

The almost frozen spectral evolution exhibited after peak by slow SLSNe~I \citep{ni16,in17} supports our findings and the idea that the photosphere reaches the inner shell earlier than in the fast events. In addition, the slow events show forbidden lines earlier, suggesting they become optically thin earlier. This could be explained by a high amount of oxygen \citep[$\sim10$\M,][]{je17}, whose recombination could hasten the process. 

Generally, differences in properties in SN subclasses, and hence between the slow and the fast, could be due to different degrees of mixing or geometry \citep{le15b,in16a,le17}, which is true regardless of the source of additional luminosity. The photospheric velocity depends on the the optical depth, for which the heavy elements with higher line opacities are the prime contributor.
A more efficient mixing of heavy elements in the outer ejecta might result in an initial higher photospheric velocity, whereas a less efficient one could lead to slow velocity and constant temperature. The gradient evolution may also be explained in terms of the ejecta density structure or of the photosphere moving to non-mixed layers of the ejecta.

\subsection{Consequences for standardization}

Figure~\ref{fig:4OPS} can be used to define a homogeneous population of events, with 90\% of previously classified SLSNe I meeting our definition, for further study in a cosmological context. This can be used for current (e.g,. DES, Pan-STARRS) and new generation (e.g., LSST, {\it Euclid} and \textit{WFIRST}) surveys to identify/classify (also in real time) SLSNe~I. This would allow identification even without a spectroscopic evolution, important given relatively limited spectroscopic resources. Moreover, with only a spectrum at +10 days, the identification can be confirmed as well as distinguishing between the fast and slow subgroups. This, together with Figure~\ref{fig:4OPS}, strengthens further the possibility to use them as standardizable candles at high-redshift. The next logical step is to discern the two subclasses only with photometry and/or to move this analysis to shorter wavelengths (e.g., the rest-frame ultraviolet) allowing higher redshifts to be studied.

\software{snake \citep{in16}, george \citep{george}}

\acknowledgments{We acknowledge support from EU/FP7-ERC grant 615929. We thank two anonymous referees, a statistician and an astronomer, for their suggestions that have improved the paper. CI thanks Stuart Sim for stimulating discussions, as well as the organisers and participants of the Munich Institute for Astro- and Particle Physics (MIAPP) workshop ``Superluminous supernovae in the next decade". }
\bibliographystyle{apj}
\bibliography{cosmo.bib}\label{bib}

\appendix
\section{Gaussian Process fits}\label{sec:GPfits}

\begin{figure*}
\center
\includegraphics[width=18cm]{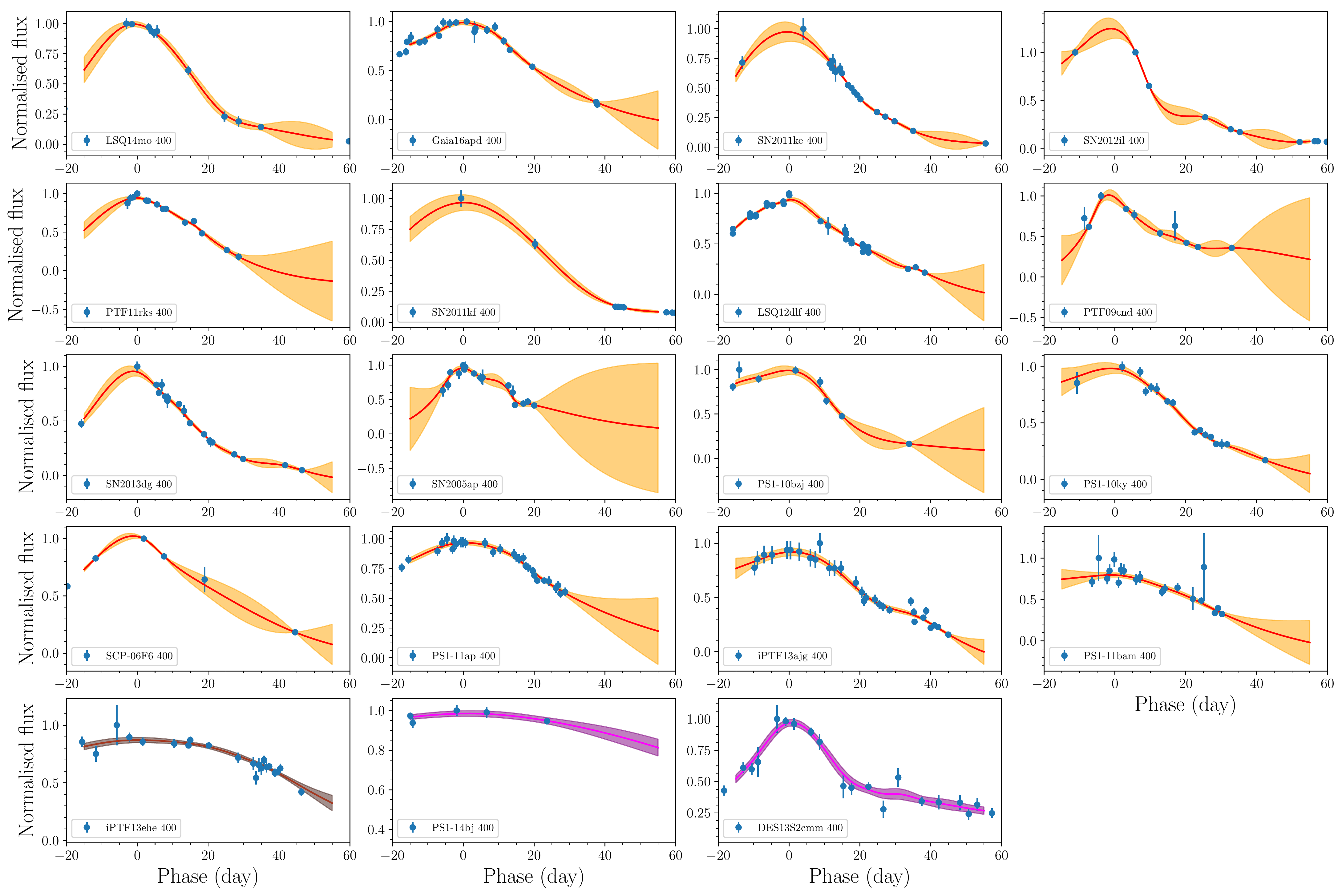}
\caption{Gaussian process fits in the 400nm band for all SLSNe listed in Table~\ref{table:data}, with the exception of those shown in Figure~\ref{fig:GP}. The test object (iPTF13ehe) and the two outliers (PS1-14bj, DES13S2cmm) are highlighted by fits of different colors. GP fits are solid lines, while the uncertainties (68\% confidence interval) are the shaded areas. }
\label{fig:GP400}
\end{figure*}

\begin{figure*}
\center
\includegraphics[width=18cm]{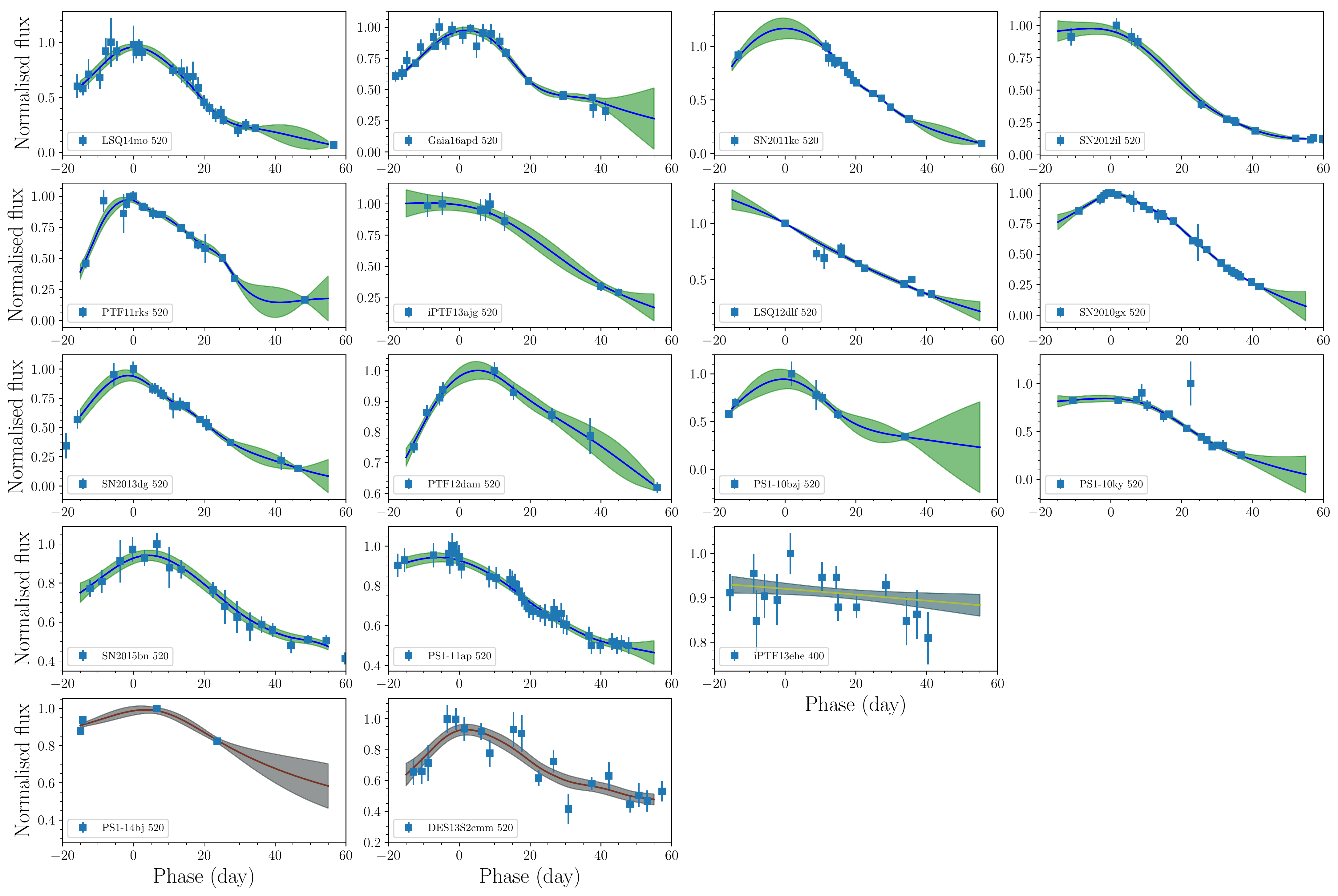}
\caption{Gaussian process fits in the 520nm band for all SLSNe listed in Table~\ref{table:data}. The test object (iPTF13ehe) and the two outliers (PS1-14bj, DES13S2cmm) are highlighted by fits of different colors.GP fits are solid lines, while the uncertainties (68\% confidence interval) are the shaded areas. The phase $t=0$ is given by the 400nm band fits in Figure~\ref{fig:GP400}. }
\label{fig:GP520}
\end{figure*}

\end{document}